\documentclass[12pt]{article}

\usepackage{color}
\usepackage{multirow}
\usepackage{graphicx}
\usepackage{hyperref}
\usepackage{amsmath}
\usepackage{amssymb}
\usepackage{amsfonts}

\advance\voffset by -2.0cm \advance\hoffset by -1.25cm
\usepackage{hyperref}

\usepackage{cite}
\usepackage{amsmath}
\usepackage{amssymb}
\usepackage{amsthm}

\usepackage{graphicx,color}

\newcommand{\RR}{\mathbb{R}}

\newcommand{\ZZ}{\mathbb{Z}}

\newcommand{\F}{\mathcal{F}}

\newcommand{\Hh}{\mathcal{H}}
\newcommand{\I}{\mathcal{I}}

\DeclareMathOperator{\Aut}{Aut}

\numberwithin{equation}{section}

\def\be{\begin{equation}}
\def\ee{\end{equation}}

\allowdisplaybreaks[3]

\begin{document}
\centerline{\large{3D String Theory and Umbral Moonshine}}
\bigskip
\bigskip
\centerline{Shamit Kachru$^1$, Natalie M. Paquette$^1$, and Roberto Volpato$^{1,2}$}
\bigskip
\bigskip
\centerline{$^1$Stanford Institute for Theoretical Physics}
\centerline{Department of Physics, Stanford University, Palo Alto, CA 94305, USA}
\medskip
\centerline{$^2$Theory Group, SLAC}
\centerline{Menlo Park, CA 94309, USA}
\bigskip
\bigskip
\begin{abstract}
The simplest string theory compactifications to 3D with 16 supercharges -- the heterotic string on $T^7$,
and type II strings on $K3 \times T^3$ -- are related by U-duality, and share a moduli space of vacua
parametrized by $O(8,24;{\mathbb Z}) ~\backslash ~O(8,24)~ \slash~ (O(8) \times O(24))$.  One can think
of this as the moduli space of even, self-dual 32-dimensional lattices with signature (8,24).  At 24 special points in
moduli space, the lattice splits as $\Gamma^{8,0} \oplus \Gamma^{0,24}$.  $\Gamma^{0,24}$ can be the 
Leech lattice or any of 23 Niemeier lattices, while $\Gamma^{8,0}$ is the $E_8$ root lattice.  We show that
starting from this observation, one can find a precise connection between the Umbral groups and type IIA string theory on $K3$.  This provides a natural physical starting point for understanding Mathieu and Umbral moonshine.  The maximal unbroken subgroups of Umbral groups in 6D (or any other limit) are those obtained by starting at the associated Niemeier point and moving in moduli space while preserving the largest possible subgroup of the Umbral group. 
To illustrate the action of these symmetries on BPS states, we discuss the computation of certain protected
four-derivative terms in the effective field theory, and recover facts about the spectrum and symmetry
representations of 1/2-BPS states.

\end{abstract}
\section{Introduction}

Since the discovery of Mathieu moonshine in the elliptic
genus of K3 \cite{EOT,Gannon}, it has been an interesting
problem to physically explain the appearance of $M_{24}$ symmetry.  It soon emerged that $M_{24}$ is not
a symmetry of any $K3$ sigma model, though the group
$Co_0$ does play a preferred role in that any automorphism group of the $K3$ CFT appears as a subgroup of $Co_0$ \cite{GHV}.

More recently, a family of 23 moonshines -- the ``Umbral
moonshines" \cite{Umbral, UMproof} -- has been described, with
the $M_{24}$ moonshine being a special case.  Umbral moonshine associates to each of the 23 Niemeier lattices -- the even, positive-definite, unimodular lattices in 24 dimensions with some vectors of ${\rm length}^2  = 2$ -- a symmetry group
(the automorphism group of the corresponding lattice), and a set of (vector-valued) mock modular forms.  
The instances of Umbral moonshine are all also conjecturally tied to K3 geometry -- see
e.g. \cite{Nikulin, CH}.

In this paper, we provide a precise connection between
the Umbral groups and string theory on K3. We observe
that each of the Umbral groups -- as well as the symmetry group 
of the Leech lattice (the 24th positive even unimodular lattice in dimension 24), $Co_0$ -- can be found in a very natural way as a symmetry of a preferred compactification of string theory to three dimensions with 16 supercharges.\footnote{The existence of the Leech lattice point in the moduli space and a possible connection with Mathieu moonshine was suggested to one of us (R.V.) by Boris Pioline.}  This compactification has two natural avatars -- via heterotic strings on $T^7$, and via type II strings on $K3 \times T^3$. We shall show that the points in moduli space which enjoy the closest connection to the Umbral groups can be described as perturbative heterotic models with generic gauge group $U(1)^{30}$. All such models, however,  are necessarily ``non-perturbative" in the type II frame.

This construction sheds direct light on various aspects of Umbral moonshine.  The Umbral groups now appear
as precise symmetries of string compactification; and upon decompactification to type II on K3 (where their existence was originally inferred), we obtain a clear picture of which subgroups should remain unbroken, at which loci in K3 moduli
space.  This should provide a physical derivation of the relevance of ``Niemeier markings'' in studies of Umbral
moonshine and K3 twining genera, as well as explaining why only 4-plane preserving subgroups of various
symmetry groups appear manifestly in the 6D limit \cite{CDHK,CFHP, DMC}.  In addition, we can directly see the contributions
of BPS states of string theory in 3D to the computation of certain ``protected" $F^4$-type terms \cite{OP}.  This
allows us to read off the degeneracies and representations of 1/2-BPS states directly from a quantity appearing in
the space-time effective action.

The organization of this note is as follows.  In \S2, we review basic facts about heterotic strings on $T^7$ / type II strings on $K3 \times T^3$, and identify 24 points of special interest in their moduli space.  In
\S3, we describe the computation of a class of $F^4$-type terms which are governed by {1/2}-BPS states in the vicinity of these special points.  In \S4, we describe some weak-coupling limits of our picture.
We conclude with a discussion
of the implications of our results for Umbral moonshine in \S5. Several technical details are relegated to the appendices.

\section{Welcome to Flatland}

\subsection{Basic facts about 3D theories with sixteen supercharges}

Here, we review the simplest story of string compactification to three dimensions preserving sixteen supercharges
\cite{Sen} (see also \cite{Marcus}).
Compactifying either the $E_8 \times E_8$ or $SO(32)$ heterotic string on $T^7$, one finds a Narain moduli space
of vacua \cite{Narain}
\begin{equation}
\label{modspace}
SO(7,23;{\mathbb Z})~\backslash ~SO(7,23) ~\slash~(SO(7) \times SO(23))
\end{equation}
given by geometric moduli, $B$-fields, and Wilson lines on the torus.

One can explicitly parametrize this space as follows.  The low-energy effective action in 3D is given by
\begin{eqnarray}
\label{action}
S &=& {1\over 4}\int~d^3x~\sqrt{-g}[ R_g - g^{\mu\nu}\partial_{\mu}\Phi \partial_{\nu}\Phi - {1\over 12}
e^{-4\Phi}g^{\mu\mu^\prime}g^{\nu\nu^\prime}g^{\rho\rho^\prime} H_{\mu\nu\rho}H_{\mu^\prime\nu^\prime\rho^\prime}\\
&-& e^{-2\Phi}g^{\mu\mu^\prime}g^{\nu\nu^\prime}F_{\mu\nu}^{(a)}(LML)_{ab} F_{\mu^\prime \nu^\prime}^{(b)} 
+ {1\over 8}g^{\mu\nu} {\rm Tr}(\partial_{\mu}M L \partial_{\nu}M L)]~.\nonumber
\end{eqnarray}
Here, $g$ is the 3D metric, $\Phi$ is the 3D heterotic dilaton, and $a = 1,\cdots,30$ parametrizes the 30 abelian gauge
fields present at generic points in the moduli space of vacua.

The moduli themselves appear in $M$, a $30\times 30$ matrix:
\begin{equation}
\left(\begin{array}{ccc}G^{-1}&G^{-1}(B+C)&G^{-1} A\\
(-B + C)G^{-1}&(G-B+C)G^{-1}(G+B+C)&(G-B+C)G^{-1}A\\
A^T G^{-1}&A^T G^{-1}(G + B + C)&I_{16} + A^T G^{-1}A\end{array}\right) ~.
\end{equation}
This matrix satisfies
\begin{equation}
\label{constraints}
MLM^T = L,~M^T = M,~L = \left(\begin{array}{ccc}0&I_7&0\\I_7&0&0\\0&0&-I_{16}\end{array}\right)~.
\end{equation}
The entries in $M$ should be thought of as follows. Let
$G_{mn}$, $B_{mn}$ and $A_m^I$ be $7 \times 7$, $7 \times 7$, and $7 \times 16$ matrices parametrizing the metric,
B-field, and Wilson lines along $T^7$ in the obvious way.  Then $C_{mn} = {1\over 2}A_m^I A_n^I$ is a $7 \times 7$ matrix as well.
Altogether, this gives 28 + 21 + 112 = 161 moduli, as is appropriate for the coset (\ref{modspace}).

The action (\ref{action}) is manifestly invariant under the transformations 
\begin{equation}
M \to \Omega M \Omega^T,~A_{\mu}^{(a)} \to \Omega_{ab}A_\mu^{(b)}
\end{equation}
with $g, B, \Phi$ invariant, as long as $\Omega$ is an $O(7,23)$ matrix satisfying
\begin{equation}
\Omega L \Omega^T = L~.
\end{equation}

However, this story can be be improved.  In three dimensions, one can dualize an abelian vector field and trade it
for a scalar.  In this case, it is useful to define the scalars via
\begin{equation}
\label{dualphotons}
\sqrt{-g} e^{-2\Phi}g^{\mu\mu^\prime}g^{\nu\nu^\prime}(ML)_{ab} F_{\mu^\prime\nu^\prime}^{(b)} =
{1\over 2}\epsilon^{\mu\nu\rho} \partial_{\rho}\psi^a~.
\end{equation}
In terms of these scalars and the original moduli, one obtains a new improved matrix of moduli,
the $32 \times 32$ matrix ${\cal M}$:

\begin{equation}
{\cal M} = \left(\begin{array}{ccc}  e^{2\Phi} & - e^{2\Phi}\psi^T & -\frac{1}{2}e^{2\Phi}\psi^T L \psi\\

-e^{2\Phi}\psi & M + e^{2\Phi}\psi \psi^T  &  ML\psi + \frac{1}{2} e^{2\Phi}\psi(\psi^T L \psi)\\

-\frac{1}{2}e^{2\Phi}\psi^T L \psi  &  \psi^T LM +\frac{1}{2}e^{2\Phi}\psi^T(\psi^T L \psi)   & 
e^{-2\Phi}+\psi^TLML\psi + {1\over 4}e^{2\Phi}(\psi^T L \psi)^2  \end{array}\right)~.
\end{equation}
This matrix satisfies
\begin{equation}
{\cal M}^T = {\cal M},~{\cal M}^T {\cal L} {\cal M} = {\cal L}
\end{equation}
with
\begin{equation}
{\cal L} = \left(\begin{array}{ccc}0&0&1\\0&L&0\\1&0&0\end{array}\right)~.
\end{equation}
It will be convenient for us later to express ${\cal M}$ in terms of a vielbein, similarly to \cite{OP},
\begin{equation}\label{vielbein}
e_{8,24} = \left(\begin{array}{ccc}g_{3H}^2&&\\
&e_{7,23}&\\&&g_{3H}^{-2}\end{array}\right) \cdot
\left(\begin{array}{ccc}1&-\psi^T&-\psi^T L \psi/2\\&
I_{30}&L\psi\\
&&1\end{array}\right),~~e^T_{8,24}\mathcal{L} e_{8,24} = \mathcal{L}~
\end{equation}
in terms of which $\mathcal{M} = e^{T}_{8, 24} e_{8, 24}$. Here, $g_{3H}^2=e^{\Phi}$ is the 3D heterotic string coupling constant, and $e_{7,23}$ is the vielbein of the perturbative heterotic moduli on $T^7$ satisfying $e^T_{7,23}L e_{7,23}=L$,  $e^T_{7,23} e_{7, 23} = M$.

In terms of these 192 scalars -- the original moduli, the dilaton, and the 30 dualized photons -- the action
simplifies to
\begin{equation}
\label{simpleac}
S = {1\over 4} \int ~d^3x~\sqrt{-g}\left( R_g + {1\over 8} g^{\mu\nu} {\rm Tr}(\partial_\mu {\cal M}
{\cal L}\partial_{\nu} {\cal M} {\cal L})\right)~.
\end{equation}
Here and in the following, we restrict ourselves to backgrounds  $H_{\mu\nu\rho}=0$.  
$S$ is now invariant under $O(8,24)$ transformations 
\begin{equation}
{\cal M} \to \Omega {\cal M} \Omega^T
\end{equation}
where $\Omega$ is any $32\times 32$ matrix satisfying
\begin{equation}
\Omega {\cal L} \Omega^T = {\cal L}~.
\end{equation}

It is a familiar fact, however, that the $O(8,24)$ group of naive symmetries will be broken to a discrete
group by non-perturbative effects (and charge quantization).
The non-perturbative duality group acting on the matrix ${\cal M}$ is in fact $O(8,24;{\mathbb Z})$.
One can roughly see this from the fact that the duality group of heterotic strings on $T^6$ is
$O(6,22;{\mathbb Z}) \times SL(2,{\mathbb Z})$ where the first factor is the T-duality group, and
the second is the famous S-duality of 4D ${\cal N}=4$ string compactifications.  We can view the 3D
theory as arising from a 4D theory in 7 different ways, as any of the circles of the $T^7$ can be made
large to give a 4D ${\cal N}=4$ limit.  This means that in addition to the $O(7,23;{\mathbb Z})$ symmetry
expected from Narain compactification to 3D as a T-duality group, we should expect 7 different ways of
finding extensions by $SL(2,{\mathbb Z})$.  The combination of these seven non-commuting S-duality groups
with $O(7,23;{\mathbb Z})$ naturally enlarges the duality group to $O(8,24;{\mathbb Z})$.  In any case,
the final non-perturbative moduli space of this connected component of the space of theories with 16 supercharges
is
\begin{equation}
\label{mspace}
O(8,24;{\mathbb Z}) ~\backslash~O(8,24)~\slash~(O(8) \times O(24))~.
\end{equation}

We close by noting that all of these facts have a dual interpretation in type II string theory on $K3 \times T^3$.
Type II strings on $K3$ and heterotic strings on $T^4$ already enjoy a string-string duality in 6D \cite{Hull}.

\subsection{Enter the Niemeier lattices}\label{s:Niemeier}

The moduli space of vacua (\ref{mspace}) can be thought of as the space of 32-dimensional even unimodular lattices with signature (8,24). These lattices can be parameterized in terms of the splitting of $\Gamma^{8,24}\otimes\RR$ into an orthogonal sum $\RR^{8}\oplus\RR^{24}$ of a positive-definite and a negative-definite subspace.  Perforce, a (non-perturbative) string vacuum in this moduli space can be specified by the
choice of such a lattice.  It is important to emphasize, however, that unlike the case with heterotic compactification
to 2D, where such a lattice would specify a point in Narain moduli space, here the lattice does not (generically) have a purely worldsheet description.  It is parametrizing a point in the full moduli space of vacua, generically with a
string coupling of ${\cal O}(1)$ in every perturbative string duality frame.

A particularly nice set of choices give lattices of the form
$$\Gamma^{8,0} \oplus \Gamma^{0,24}$$
with both factors being even unimodular lattices.  In dimension 8 there is a unique such object -- the $E_8$ lattice.
For $\Gamma^{0,24}$, however, we have 24 choices.\footnote{See
e.g. \cite{Conway} for a discussion of even unimodular lattices of definite signature in various dimensions.}

\medskip
\noindent
$\bullet$ We can choose the Leech lattice, the unique
positive-definite even unimodular lattice of rank 24 with no roots.  It has a symmetry group $Co_0$, closely related
to the sporadic simple group $Co_1$.

\medskip
\noindent
$\bullet$ There are also 23 Niemeier lattices $L^X$, associated to the 23 Niemeier root systems $X$. The root systems $X$ are unions of A-D-E root systems that have total rank 24 and share a Coxeter number.  We list the Niemeier lattices (labelled by the A-D-E root systems of which they are comprised), as well as the associated Umbral symmetry
groups, in the table below.  The Umbral symmetry group should be thought of as the automorphism group of the lattice
${\rm Aut}(L^X)$, modulo any Weyl group elements.  It is important to recall that the Niemeier lattices  contain additional ${\it gluing~vectors}$ in addition to the lattice spanned by the A-D-E root system, so the symmetry groups are not completely obvious given the root system.

$$\begin{array}{cc}
{\bf Niemeier ~root ~system ~X}&{\bf Umbral ~group~ G^X}\\
A_1^{24}&M_{24}\\
A_2^{12}&2.M_{12}\\
A_3^8&2.AGL_3(2)\\
A_4^6&GL_2(5)/2\\
A_5^4D_4&GL_2(3)\\
A_6^4&SL_2(3)\\
A_7^2D_5^2&{\rm Dih}_4\\
A_8^3&{\rm Dih}_6\\
A_9^2D_6&{\mathbb Z}_4\\
A_{11}D_7E_6&{\mathbb Z}_2\\
A_{12}^2&{\mathbb Z}_4\\
A_{15}D_9&{\mathbb Z}_2\\
A_{17}E_7&{\mathbb Z}_2\\
A_{24}&{\mathbb Z}_2\\
D_4^6&3.{\rm Sym}_6\\
D_6^4&{\rm Sym}_4\\
D_8^3&{\rm Sym}_3\\
D_{10}E_7^2&{\mathbb Z}_2\\
D_{12}^2&{\mathbb Z}_2\\
D_{16}E_8& 1\\
D_{24}& 1\\
E_6^4&GL_2(3)\\
E_8^3&{\rm Sym}_3
\end{array}$$

When we choose $\Gamma^{0,24} = L^X(-1)$ for some $X$,\footnote{Here and in the following, $L(-1)$ denotes a lattice $L$ where the sign of the quadratic form is flipped.} we find a theory whose symmetry group is the product of the Umbral group $G^X$ and a continuous non-abelian gauge symmetry, whose Lie algebra is specified by the root system $X$. Due to this enhanced gauge symmetry, these theories cannot be described by the simple effective action \eqref{simpleac}. Furthermore, various BPS-saturated threshold corrections (including the ones we will consider in the next sections) diverge at these enhanced gauge symmetry points. We will show that there exist deformations of the moduli that break the enhanced gauge group to the generic $U(1)^{30}$ while preserving the whole finite symmetry group $G^X$. In fact, $G^X$ is the maximal subgroup that can be preserved for a model with generic gauge group $U(1)^{30}$, at least in a neighborhood of a `Niemeier  point' in the moduli space.

\begin{figure}\label{figure}
\centering
\includegraphics[scale=0.33]{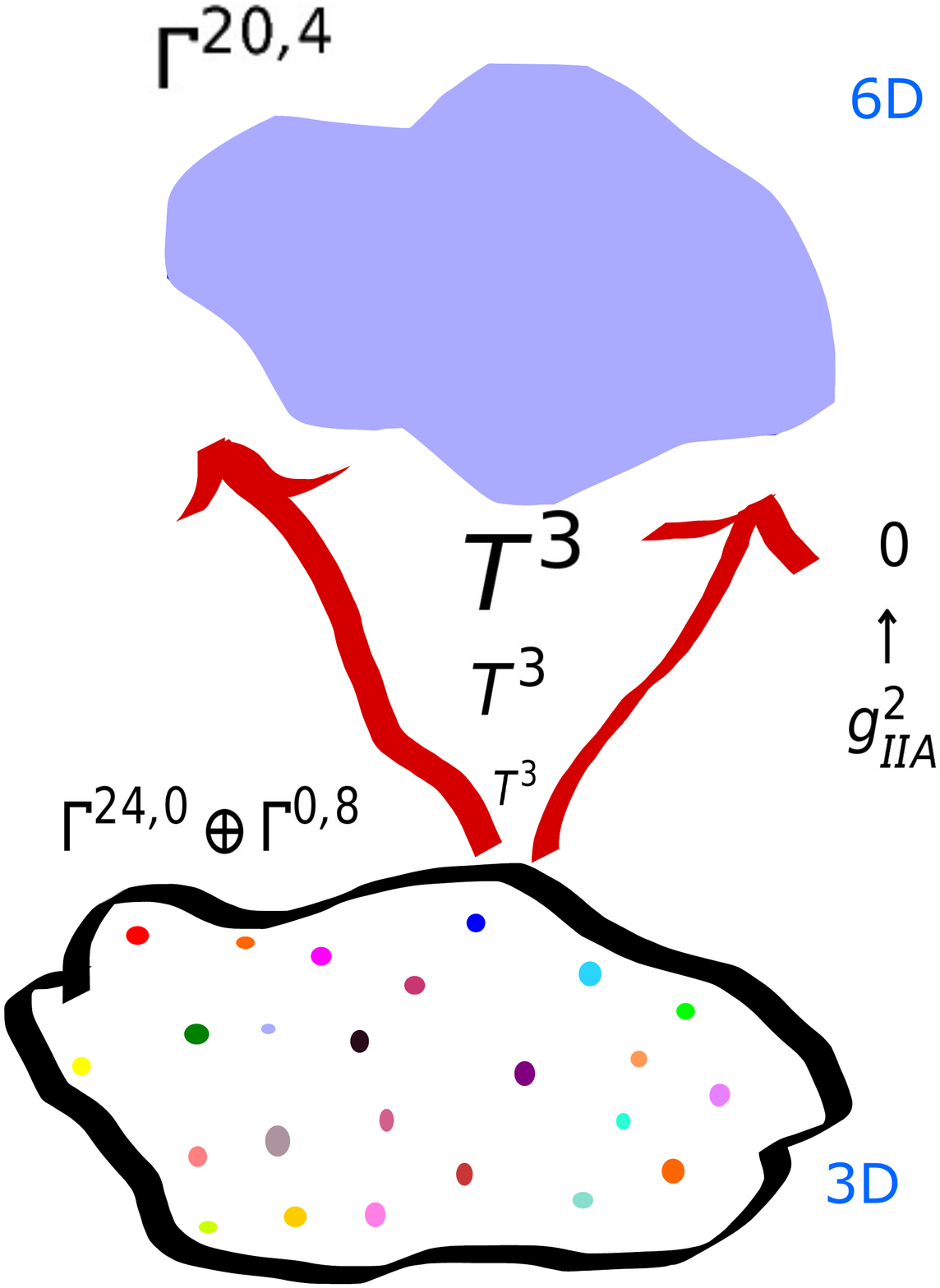}
\caption{The Umbral groups arise as symmetries of string theory near the 23 Niemeier points in the moduli
space of 3D vacua.  Decompactification to 6D breaks (some of) the Umbral symmetry, leaving the groups known to appear in K3 compactification of IIA string theory.}
\end{figure}

This means that
each Umbral symmetry
group $G^X$ arises as the symmetry group of the low energy effective action \eqref{simpleac} (which extends to a symmetry of the full
high energy theory).  In particular, starting from $\Gamma^{0,24} = L^{A_{1}^{24}}(-1)$, we will find a theory on
the moduli space of type II compactifications on $K3 \times T^3$, with full $M_{24}$ symmetry.
It is not a perturbative string theory.\footnote{Though one could find a perturbative description upon
further reduction to 2D, if one desired.} Our thesis is that this $M_{24}$ explains the appearance of
$M_{24}$ representations in type IIA compactifications on $K3$, and the fact that one must `decompactify'
to get to the 6D limit (breaking some of the symmetry, as we shall discuss in \S4), is behind the need
to study only proper subgroups in 6D.  For instance, choosing instead the Leech lattice, we would find a 
$Co_0$ symmetry in 3D; but the need to choose a decompactification limit to 6D would reduce manifest symmetries
in the 6D theory to subgroups of $Co_0$ that preserve a 4-plane in the ${\bf 24}$, as described in 
\cite{CDHK}.  The 3D origin of the Umbral symmetries therefore dovetails nicely with the higher-dimensional
description, which sees only symmetries that are preserved in a given choice of the decompactification limit from
3D to 6D. Our philosophy is encapsulated in Figure 1.

\section{Threshold corrections in 3D theories}

Although the discussion thus far already makes it clear that Umbral symmetry groups arise in the low-energy physics, 
and hence will act non-trivially on the BPS states of type II string compactifications on $K3 \times T^3$, it
is nice to add some more flesh to the discussion.  Therefore, we discuss a class of amplitudes which are saturated
by the contributions of BPS states.  These are terms that were studied in the earlier literature by Obers and
Pioline \cite{OP}.  

\subsection{Basic facts}

The two-derivative effective action in theories with 16 supercharges does not receive interesting quantum corrections.
However, there are four-derivative terms which do receive corrections, entirely from BPS states.  In higher
dimensions, these terms would be of the schematic form
$${\cal L} \supset f(\phi) {\rm Tr}(F^4)$$
and the function of moduli $f(\phi)$ would be the object of interest.  In 3D, at generic points in moduli space where the gauge group is a product of abelian factors, one can dualize all 30 of the gauge fields to scalars. The U-duality group should mix the scalars $\phi$ amongst themselves, so we are really interested in terms of the schematic form
$$f(\phi) (\partial\phi)^4~.$$

Starting with the 1-loop computation in perturbative heterotic string theory and promoting it to a U-duality invariant
expression, Obers and Pioline were led to propose that these terms take the form
\be\label{thresh} I_{(\partial \phi)^4} ~=~g_{3H}^2 l_H \int d^3x~\sqrt{g} \int_{\cal F} {d^2\tau \over \tau_2^2} ~
{Z_{8,24}(G/l_H^2,B,\psi,g_{3H}^2) \over \eta^{24}}~ t_8[(e^{-1}_{8,24}\partial_{\mu}e_{8,24})_{ia}p_L^a]^4~.\ee
Very roughly speaking, one should think of the $(e^{-1}_{8,24}\partial_{\mu}e_{8,24})_{ia}$ term, as a particular component of $(\partial \phi)^4$, for some of the moduli $\phi$; more formally, it is (the pullback to spacetime of) a left-invariant one-form on the coset space $(SO(8)\times SO(24)) \backslash SO(8, 24)$. $t_8$ is a standard eight-tensor which is given explicitly in, e.g., appendix 9A of \cite{GSW}. Finally, $Z_{8,24}$ is the theta function of the non-perturbative $\Gamma^{8,24}$ lattice specifying the point in
moduli space where we are computing the coupling. In more explicit terms, it takes the form
\begin{equation}
Z_{8,24}(\tau,\bar\tau) = (\tau_2)^4 \sum_{v\in \Gamma^{8,24}} {\rm Exp} \left({-\pi \tau_2 v^T \mathcal{M} v + \pi i \tau_1 v^T \mathcal{L} v}\right) =(\tau_2)^4 \sum_{(p_L,p_R)\in \Gamma^{8,24}}q^{\frac{p_L^2}{2}}\bar q^{\frac{p_R^2}{2}}~.
\end{equation}
Here, $v$ is the vector of momenta, windings, and gauge charges, taking values in the even unimodular lattice $\Gamma^{8,24}$ with quadratic form $\mathcal{L}$.
We also defined the left- and right-moving momenta
\be p_L=\frac{\mathbf{1}_{32}-\mathcal{L}}{2}e_{8,24} v,\qquad p_R=\frac{\mathbf{1}_{32}+\mathcal{L}}{2}e_{8,24} v,\qquad\quad v\in \Gamma^{8,24}~.
\ee
Our convention will be the ``left'' or ``holomorphic'' side is the 24-dimensional side, so that a vector $(p_L,p_R)\in \Gamma^{8,24}$ has norm $p_R^2-p_L^2$. Expanding about a Niemeier point, $p_L^a$ is a vector in the Niemeier lattice.

We can unpackage this formula a little bit as follows.  The $p^4$ insertion in the integral is a schematic notation
for a set of derivatives that has nice modular properties, c.f. appendix \ref{app}.  Choosing a single vector in the left-moving lattice we can even more explicitly write the integral we need to evaluate to compute the four-derivative coupling in the space-time
action, as
\begin{equation}
\label{pigpigpig}
I = \int_{\cal F} {d^2\tau \over \tau_2^2} ~\tau_2^4 \left( \sum_{(p_L,p_R) \in \Gamma^{8,24}} 4 (p_{L,1}^4 - {6\over 4\pi \tau_2}p_{L,1}^2
+ {3\over 16\pi^2\tau_2^2})q^{p_L^2 \over 2} \bar q^{p_R^2 \over 2}\right) ~{1\over \eta^{24}(\tau)}~,
\end{equation}
where $p_L=(p_{L,1},\ldots,p_{L,24})$. It is this expression that we will evaluate.

\subsection{Evaluating the integral near Niemeier lattice points in moduli space}\label{Evaluation}


Using the techniques for evaluating integrals like (\ref{pigpigpig}) developed in \cite{Pioline} and references
therein (see appendix \ref{app}), we find

\be\label{integr} I=\lim_{t\to 0}\sum_{\substack{(p_L,p_R)\in \Gamma^{8,24}\\ p_R^2-p_L^2=-2}} G(p_L^2,p_{L,1}^2,t)
\ee where
\begin{align}\nonumber
G(p_L^2,p_{L,1}^2,t)=\frac{4}{13!}\frac{1}{(4 \pi)^3} & \Bigl( \Gamma(16+t) p_{L, 1}^{4} { \ }_2 F_{1}\left(1+t, 16+t, 14+2t, \frac{2}{p_L^2} \right) (p_L^2/2)^{-16-t} \\ \nonumber & + 6\Gamma(15+t) p_{L, 1}^{2} { \ }_2 F_{1}\left(1+t, 15+t, 14+2t, \frac{2}{p_L^2} \right) (p_L^2/2)^{-15-t}
\\ \nonumber &+ 3\Gamma(14+t) { \ }_2 F_{1}\left(1+t, 14+t, 14+2t, \frac{2}{p_L^2} \right) (p_L^2/2)^{-14-t} \Bigr)
\end{align}
Unfortunately, one cannot simply evaluate $I$ by replacing the value $t=0$ in each terms of the right-hand side of \eqref{integr}, because the sum only converges for $\Re{t}>1$. The right-hand side of \eqref{integr} should be really interpreted as the analytic continuation to $t=0$ of the function $I(t)$ which, for $\Re{t}>1$, is defined by the series on the right-hand side.

In any case, it is useful to study the general term of the sum for $t=0$,
\begin{align}\nonumber
G(p_L^2,p_{L,1}^2,0)=\frac{4}{13!}\frac{1}{(4 \pi)^3} & \Bigl( \Gamma(16) p_{L, 1}^{4} { \ }_2 F_{1}\left(1, 16, 14, \frac{2}{p_L^2} \right) (p_L^2/2)^{-16} \\ \nonumber & + 6\Gamma(15) p_{L, 1}^{2} { \ }_2 F_{1}\left(1, 15, 14, \frac{2}{p_L^2} \right) (p_L^2/2)^{-15}
\\ \nonumber &+ 3\Gamma(14) { \ }_2 F_{1}\left(1, 14, 14, \frac{2}{p_L^2} \right) (p_L^2/2)^{-14} \Bigr)
\end{align}
The hypergeometric functions are given by
\begin{align}
&{ \ }_2 F_{1}\left(1, 16, 14, z \right)=(1-z)^{-3}{ \ }_2 F_{1}\left(13, -2, 14, z \right)=
(1-z)^{-3}\left(1-\frac{13}{7}z+\frac{13}{15}z^2\right)\\
&{ \ }_2 F_{1}\left(1, 15, 14, z \right)=(1-z)^{-2}{ \ }_2 F_{1}\left(13, -1, 14, z \right)=
(1-z)^{-2}\left(1-\frac{13}{14}z\right)\\
&{ \ }_2 F_{1}\left(1, 14, 14, z \right)=\frac{1}{1-z}
\end{align} where we used the identity
\be { \ }_2 F_{1}\left(a, b; c; z \right)=(1-z)^{c-a-b}{ \ }_2 F_{1}\left(c-a, c-b; c; z \right)\ .
\ee We notice immediately that the sum diverges whenever there are momenta $(p_L,p_R)$ with $p_L^2=2$ and $p_R^2=0$, i.e. at the points of enhanced gauge symmetries.

Let us now specialize this result to the points in the moduli space where $\Gamma^{8,24}\cong E_8\oplus L(-1)$, where $L$ is a positive-definite, even, unimodular lattice.  For each such  lattice $L$ and vector $\lambda\in L$, we consider the theta series
\be \Theta_{L,\lambda}(\tau,z)=\sum_{\lambda'\in L} q^{\frac{(\lambda')^2}{2}}y^{\lambda'\cdot \lambda}= \sum_{n,r\in \ZZ} c_{L,\lambda}(n,r) q^ny^r\ ,
\ee which is a Jacobi form of weight $w=\frac{1}{2}\dim(L)$ and index $m=\frac{\lambda^2}{2}$ \cite{EichlerZagier}. The Fourier coefficients $c_{L,\lambda}$ correspond to
\be c_{L,\lambda}(n,r)=\#\{\lambda'\in L\mid \lambda'^2=2n,\ \lambda\cdot\lambda'=r\}\ .
\ee Notice that since $(\lambda'\cdot\lambda)^2\le \lambda^2{\lambda'}^2$, one has $c_{L,\lambda}(n,r)=0$ whenever $r^2 > 4mn$, so that $\Theta_{L,\lambda}(\tau,z)$ is actually holomorphic, rather than weak. For $z=0$, we recover the standard theta series
\be \Theta_L(\tau)=\Theta_{L,\lambda}(\tau,0)=\sum_{\lambda'\in L} q^{\frac{(\lambda')^2}{2}}= \sum_{n\in \ZZ} c_{L}(n) q^n\ ,
\ee where $c_L(n)=\sum_{r\in\ZZ} c_{L,\lambda}(n,r)$.

When $L$ is the Leech lattice and we take $p_{L, 1}$ to be the component of $p_L$ in the direction of a vector $\lambda\in L$, the integral becomes
\be
I=\lim_{t\to 0}\sum_{n=2}^\infty \sum_{r\in \ZZ} c_{Leech,\lambda}(n,r)c_{E_8}(n-1) G(2n,\frac{r^2}{\lambda^2},t).
\ee Each term in the sum is finite as $t\to 0$, since $G(x,y,0)$ only diverges for $x=2$ and the Leech lattice has no roots. We stress again that the left-hand side should be interpreted as the analytic continuation to $t=0$ of the series on the right-hand side, which converges only for $\Re{t}>1$.

When $L$ is a Niemeier lattice with roots, then we have to take care of the divergence of $G$ due to states with $p_L^2=2$, $p_R^2=0$. We regularize the sum by taking an infinitesimal deformation along a null direction in the moduli space.
To be explicit, let us define two orthonormal bases of vectors $e_i\in L\otimes \RR$, $i=1,\ldots,24$, and $\tilde e_j\in E_8\otimes \RR$, $j=1,\ldots,8$. In the undeformed model (i.e. at the point of enhanced gauge symmetry), the left- and right-moving momenta have components
\be p_{L, i}(\lambda,\mu)=\lambda\cdot e_i\ ,\qquad p_{R, j}(\lambda,\mu)=\mu\cdot \tilde e_j\ ,\qquad \lambda\in L\ ,\mu\in E_8 \qquad
\ee

Now we deform the moduli in a null direction along the plane spanned by $e_1,\tilde e_1$ so that the new components in the $1$-direction are given by
\be \begin{pmatrix}
p_{L,1}\\ p_{R,1}
\end{pmatrix} =\begin{pmatrix}
\cosh \epsilon & \sinh \epsilon\\ \sinh \epsilon & \cosh \epsilon
\end{pmatrix}\begin{pmatrix}
\lambda\cdot e_1 \\ \mu\cdot \tilde e_1
\end{pmatrix}\qquad \lambda\in L,\ \mu\in E_8\ ,
\ee for some small real parameter $\epsilon$; the other components $p_{L, i},p_{R, j}$, $i,j>1$, are unchanged. Notice that even in the deformed theory the momenta satisfy
\be p_{R}^2-p_{L}^2=\mu^2-\lambda^2\in 2\ZZ\ .
\ee
Choosing $e_1=\frac{\lambda}{|\lambda|}$, where $\lambda \in L$, and $\tilde e_1=\frac{\mu}{|\mu|}$, for some $\mu\in E_8$,
the deformed momenta are
\be p_{L,1}^2(\lambda',\mu')=\cosh^2\epsilon \frac{(\lambda'\cdot\lambda)^2}{\lambda^2}+2\cosh\epsilon\sinh \epsilon \frac{(\lambda'\cdot\lambda)}{|\lambda|}\frac{(\mu'\cdot\mu)}{|\mu|}+\sinh^2\epsilon\frac{(\mu'\cdot\mu)^2}{\mu^2}
\ee
\be
p_L^2(\lambda',\mu')={\lambda'}^2+\sinh^2\epsilon \frac{(\lambda'\cdot\lambda)^2}{\lambda^2}+2\cosh\epsilon\sinh \epsilon \frac{(\lambda'\cdot\lambda)}{|\lambda|}\frac{(\mu'\cdot\mu)}{|\mu|}+\sinh^2\epsilon\frac{(\mu'\cdot\mu)^2}{\mu^2} \ee
By specializing these formulae to the case where ${\lambda'}^2=2$ and $\mu'=0$, we get
\be p_{L,1}^2(\lambda',0)=\frac{(\lambda'\cdot\lambda)^2}{\lambda^2}\cosh^2\epsilon
\ee
\be
p_L^2(\lambda',0)=2+\frac{(\lambda'\cdot\lambda)^2}{\lambda^2}\sinh^2\epsilon \ee
and
\begin{align} I=&\sum_{r\in\ZZ}c_{L,\lambda}(1,r)G\left(2+\sinh^2\epsilon \frac{r^2}{\lambda^2},\cosh^2\epsilon\frac{r^2}{\lambda^2},0\right)\\
&+ \lim_{t\to 0}\sum_{n=2}^\infty \sum_{r\in \ZZ} c_{L,\lambda}(n,r)c_{E_8}(n-1)
G(2n,\frac{r^2}{\lambda^2},t)+O(\epsilon^2)\\
=&\sum_{r\in\ZZ}c_{L,\lambda}(1,r)\frac{4}{(4\pi)^3}\left(\frac{16}{r^2/\lambda^2 }\epsilon^{-6}+\frac{40}{r^2/\lambda^2 }\epsilon^{-4}+(\frac{56}{5r^2/\lambda^2 }+6\frac{r^2}{\lambda^2})\epsilon^{-2}
\right.\\&\qquad\qquad\qquad\qquad\quad \left.
-\frac{544}{189 r^2/\lambda^2 }-536\frac{r^2}{\lambda^2}-949 \frac{r^4}{\lambda^4} \right)\\
&+ \lim_{t\to 0}\sum_{n=2}^\infty \sum_{r\in \ZZ} c_{L,\lambda}(n,r)c_{E_8}(n-1)
G(2n,\frac{r^2}{\lambda^2},t)+O(\epsilon^2)
\end{align}
Notice that the dependence on $\mu$ (i.e. on the direction of the deformation in the $E_8\otimes \RR$ space) only appears at $O(\epsilon^2)$. The $n=1$ term is convergent for $\epsilon\neq 0$, provided that
\be c_{L,\lambda}(1,0)=0\ .
\ee This means that $\lambda$ cannot be orthogonal to any root.

\bigskip

The choice of $\lambda$ breaks the group of automorphisms of the lattice $L$ down to the subgroup $H_\lambda\subseteq \Aut(L)$ fixing $\lambda$. The group of automorphisms of $L^X$ has the form $\Aut(L^X)=W\rtimes G^X$ where $W$ is the Weyl group and $G^X$ is the Umbral group, as in section \ref{s:Niemeier}. The condition that $\lambda$ is not orthogonal to any root implies that $\lambda$ is in the interior of a Weyl chamber. Therefore, the group $H_\lambda$  is contained in the subgroup of $\Aut(L^X)$ preserving the given Weyl chamber as a set. The latter group is isomorphic to $G^X$, so that, in general, $H_\lambda\subseteq G^X$. In particular, if we choose $\lambda$ to be (proportional to) the Weyl vector relative to a given Weyl chamber, then the group fixing $\lambda$ is isomorphic to the whole $G^X$, $H_\lambda\cong G^X$. We conclude that the Umbral group $G^X$ is the maximal subgroup of $\Aut(L^X)$ that can be preserved by an infinitesimal deformation of the model to a point with generic gauge group $U(1)^{30}$.

In appendix \ref{a124example} we  give a more detailed description of the example of the Niemeier lattice with root system $X=A_1^{24}$ and the vector $\lambda$ equal to a Weyl vector. We present a decomposition of the first few coefficients $c(n, r)$ into $G^X=M_{24}$ irreps: 
\begin{align*}
c(1,1)&=24={\bf 1}+{\bf 23}\\
c(2,4)&=759={\bf 1} + {\bf 23} + {\bf 252} + {\bf 483}\\ 
c(2,3)&=6072= {\bf 1} + 2\times {\bf 23} + 2\times {\bf 252} +{\bf 253}  + {\bf 483}+{\bf 1265}+{\bf 3520} \\
&\vdots
\end{align*}

%

\section{Weak coupling limits}

In the previous section, we showed that there are 3D models with generic gauge group $U(1)^{30}$ whose group of symmetries is one of the Umbral groups $G^X$. These models are infinitesimally close to the Niemeier points in the moduli space, and as such they are intrinsically strongly coupled.

In this section, we shall deform the models to reach the weak coupling limits in the heterotic and in the type IIA frames. In general, these deformations will break the symmetry groups of the model. The idea is to choose the deformations so as to preserve the largest possible groups. Our main goal is to understand the action of the residual groups on the BPS states of the weakly coupled theory.

\subsection{Perturbative heterotic limit}\label{s:pertHet}

Let us first consider a deformation to the weak coupling limit of the heterotic string on $T^7$. There are many possible ways of taking this limit, corresponding to the choice of a splitting of the lattice $\Gamma^{8,24}\cong E_8\oplus L(-1)$ into a sum $\Gamma^{8,24}=\Gamma^{1,1}\oplus \Gamma^{7,23}$. The summand $\Gamma^{7,23}$ will be interpreted as the Narain lattice in the perturbative limit. 

 To be explicit, choose bases $\lambda_1,\ldots,\lambda_{24}$ of $L$ and $\mu_1,\ldots,\mu_8$ of $E_8$ and let $\mu_1^*,\ldots,\mu_8^*$ be the dual basis of $E_8$, satisfying $\mu_i\cdot \mu_j^*=\delta_{ij}$. For the sake of simplicity, we also assume that
\be \lambda_1^2=(\mu_1^*)^2\ .
\ee For any choice of $\lambda_1,\ldots,\lambda_{24}$, this can be always satisfied by a suitable choice of $\mu_1,\ldots,\mu_8$. Then, we define
\be\label{uvdef} u:=\lambda_1+\mu_1^*\qquad v:= -\mu_1+\frac{\mu_1^2}{2}u\ ,
\ee and
\be \lambda_i'=\lambda_i-\lambda_i\cdot\lambda_1 (u+\frac{\mu_1^2}{2}v),\qquad i=2,\ldots,24\ ,
\ee
\be \mu_i'=\mu_i+\mu_i\cdot \mu_1 u\qquad i=2,\ldots,8\ .
\ee Then, $u,v$ generate $\Gamma^{1,1}$ with standard quadratic form $\left(\begin{smallmatrix}
0 & 1\\ 1 & 0
\end{smallmatrix}\right)$ and $\lambda'_2,\ldots,\lambda'_{24},\mu_2',\ldots,\mu_8'$ generate a unimodular lattice $\Gamma^{7,23}$ orthogonal to $\Gamma^{1,1}$. We denote by $Q_{7,23}$ the quadratic form of signature $(7,23)$ of the lattice $\Gamma^{7,23}$.

Let us now consider the heterotic string frame where $\Gamma^{7,23}$ is the Narain lattice of winding and momenta.  The most general vielbein parametrizing cosets in $SO(8,24)/(SO(8)\times SO(24))$ can be written, generalizing \eqref{vielbein}, as
\be\label{vielbein2} e_{8,24}=\begin{pmatrix}
g^2_{3H} && \\ & e_{7,23} & \\ && g^{-2}_{3H}
\end{pmatrix}\begin{pmatrix}
1 & -\psi^T & -\frac{1}{2}\psi^T Q_{7,23}^{-1} \psi\\
0 & \mathbf{1}_{30} & Q_{7,23}^{-1}\psi\\
&& 1
\end{pmatrix}\ ,
\ee where $g_{3H}$ is the three dimensional heterotic string coupling constant, $e_{7,23}$ is the vielbein parametrizing the Narain moduli space $SO(7,23)/(SO(7)\times SO(23))$ and satisfying
\be e_{7,23}^T Q_{7,23} e_{7,23}=Q_{7,23}\ ,
\ee and $\psi^T:=(\psi_1,\ldots,\psi_{30})$ are the vevs of the scalars obtained by dualizing the $30$ gauge fields in three dimensions.

Let us start with the values of the moduli $g_{3H}, \psi, e_{7,23}$ corresponding to one of the Niemeier points in the moduli space, and then send $g_{3H}\to 0$ while keeping the other moduli fixed. We will prove that the perturbative heterotic string obtained in this way has a discrete symmetry group $G$ isomorphic to the subgroup of $\Aut(L)$ preserving the vector $\lambda_1\in L$. Furthermore, the symmetry acts trivially on the right-moving (supersymmetric) string oscillators, while the $23$ left-moving (bosonic) oscillators in the internal directions form a $23$ dimensional representation of $G$.

In general, the symmetries of a model at a given point $e_{8,24}$ in the moduli space are really dualities that fix that point, i.e. $h\in O(\Gamma^{28,4})$ such that
\be h\cdot e_{8,24}=e_{8,24}\cdot \rho\ ,
\ee where $\rho\in SO(8)\times SO(24)$, so that $e_{8,24}$ and $h\cdot e_{8,24}$ denote the same coset in $SO(8,24)/(SO(8)\times SO(24))$. Furthermore, if $\rho\in SO(24)$ (respectively, $\rho\in SO(8)$) then the symmetry acts non-trivially only on the left-moving (resp., right-moving) oscillators.

In the case where $e_{8,24}$ is one of the Niemeier lattice points, any $h\in \Aut(L)$ is a symmetry and the corresponding $\rho$ is contained in $SO(24)\cong SO(L\otimes\RR)$. Let us consider symmetries $h$ in the subgroup $H\subseteq \Aut(L)$  fixing the vector $\lambda_1$ in \eqref{uvdef}. Such an $h$ fixes both $u$ and $v$, so that it only acts non-trivially on its orthogonal complement $\Gamma^{7,23}$. Thus, $h$ must satisfy
\be\label{hetsymm} h\cdot e_{8,24}=\begin{pmatrix}
1 & &\\ & h' &\\ && 1
\end{pmatrix}\cdot e_{8,24}=e_{8,24}\cdot\begin{pmatrix}
1 & &\\ & \rho' &\\ && 1
\end{pmatrix}\ ,
\ee where $h'\in O(\Gamma^{7,23})$ and $\rho'\in SO(23)\subset SO(7)\times SO(23)$, where $SO(23)$ acts on the orthogonal complement of $\lambda_1$ in $L\otimes \RR$. By plugging the expression \eqref{vielbein2} into $e_{8,24}$, it is then clear that \eqref{hetsymm} holds independently of $g_{3H}$. Therefore, $h$ must be a symmetry also of the weakly coupled model and it only acts non-trivially on the left-moving bosonic oscillators.

\bigskip

When $L$ is one of the $23$ Niemeier lattices with roots, the analysis of the unbroken group of symmetries $H$ is exactly as in section \ref{Evaluation}. We can choose $\lambda_1$ to be the Weyl vector of the corresponding root lattice (for some choice of positive roots). In this case, the perturbative heterotic string has generic gauge group $U(1)^{30}$ (no enhanced gauge symmetry) and the unbroken group of symmetries is the Umbral group associated with the root lattice $X$, $H\cong G^X$. For all the other choices of $\lambda$ leading to a theory with generic gauge group $U(1)^{30}$, the symmetry group is a subgroup of the Umbral group.

When $L$ is the Leech lattice, $H$ is any subgroup of the Conway group $Co_0$ fixing a one-dimensional subspace of $L\otimes \RR$. Some interesting choices lead to $H$ being the Mathieu group $M_{24}$ or the Conway groups $Co_2$ or $Co_3$.

\subsection{Perturbative type IIA limit}

Analogous considerations hold for deformations of the Niemeier models leading to a perturbative type IIA limit. As a first step, we split the lattice $\Gamma^{8,24}$ into the orthogonal sum $\Gamma^{4,4}\oplus \Gamma^{4,20}$. Roughly speaking, this corresponds to writing the non-perturbative torus $T^8$ of the heterotic string (combining the geometric $T^7$ torus and the non-perturbatively generated circle of radius $1/g_{3H}^2$) as a product $T^8=T^4\times \tilde T^4$. Here, the non-perturbative torus $T^4$ is associated with the $\Gamma^{4,4}$ summand in $\Gamma^{8,24}$, while the $\Gamma^{4,20}$ summand is interpreted as the Narain lattice for heterotic strings on the geometric $\tilde T^4$. 

Explicitly, the lattice splitting can be done as follows. Choose, as above, bases $\lambda_1,\ldots,\lambda_{24}$ of $L$ and $\mu_1,\ldots,\mu_8$ of $E_8$, with $\mu_1^*,\ldots,\mu_8^*$ be the dual basis of $E_8$. We also assume that
\be \lambda_i\cdot \lambda_j=\mu_i^*\cdot \mu^*_j\qquad\qquad i,j=1,\ldots,4\ .
\ee For any choice of $\lambda_1,\ldots,\lambda_{24}$, there is always a basis of $E_8$ for which this is true.\footnote{This is equivalent to the fact that every even positive lattice of rank at most $4$ can be primitively embedded in $E_8$; see \cite{Nikulin} for a proof.} Then, the lattice $\Gamma^{4,4}$ is generated by
\be u_i=\lambda_i+\mu_i^* \qquad v_i=-\mu_i+\frac{\mu_i^2}{2}u_i+\sum_{k=i+1}^4 \mu_i\cdot \mu_ku_k\qquad i=1,\ldots,4
\ee and the orthogonal complement $\Gamma^{4,20}$ by
\be \lambda_i'=\lambda_i-\sum_{j=1}^4\lambda_i\cdot\lambda_j (v_j+\frac{\mu_j^2}{2}u_j)-\sum_{\substack{j,k=1\\j<k}}^4(\lambda_i\cdot\lambda_k)(\mu_j\cdot \mu_k) u_j\ ,\qquad i=5,\ldots,24\ ,\ee
\be \mu_i'=\mu_i+\sum_{k=1}^4\mu_i\cdot\mu_k u_k\ ,\qquad i=5,\ldots,8\ .
\ee We denote by $Q_{4,20}$ the quadratic form of the lattice $\Gamma^{4,20}$.

Given a decomposition of the non-perturbative heterotic $T^8$ into $T^4\times \tilde T^4$, it is useful to take a metric 
\be \mathcal{L}'=\begin{pmatrix}
&& \mathbf{1}_4\\
& Q_{4,20} &\\
\mathbf{1}_4 & &
\end{pmatrix}
\ee for the lattice $\Gamma^{4,4}\oplus \Gamma^{4,20}$.
 The vielbein parametrizing cosets in $SO(8,24)/(SO(8)\times SO(24))$ and satisfying $e_{8,24}^T\mathcal{L}'e_{8,24}=\mathcal{L}'$ can be written as \cite{OP} \be\label{IIAvielbein} e_{8,24}=\begin{pmatrix}
v^{-T} && \\ & e_{4,20} & \\ && v
\end{pmatrix}\begin{pmatrix}
\mathbf{1}_4 & -\psi^T & B-\frac{1}{2}\psi^T Q_{4,20}^{-1} \psi\\
0 & \mathbf{1}_{24} & Q_{4,20}^{-1}\psi\\
&& \mathbf{1}_4
\end{pmatrix}\ ,
\ee where, in heterotic string frame, $v$ is the vierbein of the non-perturbative torus $T^4$ (including a $T^3$ space-time torus and the dynamically generated circle of radius $1/g_{3H}^2$), whose metric is given by $G=l_H^2 v^Tv$, with $l_H$ the heterotic string length. Furthermore, $B$ is the B-field along the non-perturbative $T^4$, $\psi$ is a $(24\times 4)$-dimensional matrix of Wilson lines and $e_{4,20}$ is the vielbein parametrizing the Narain moduli space for heterotic strings on $\tilde T^4$. In the language of type IIA on $K3\times T^3$, the three-dimensional effective string coupling constant $g_{3IIA}$ is related to the volume of the non-perturbative $T^4$ torus in units of heterotic strings length \cite{OP}, i.e.\footnote{Recall that, in six dimensions, the coupling constants and string lengths of type IIA on K3 and heterotic on $T^4$ are related by $g_{6IIA}=1/g_{6H}$ and $l_H=g_{6IIA}l_{IIA}$.}
\be 1/g_{3IIA}^2=\left|\det v\right|^2\ ,
\ee so that the perturbative type IIA limit corresponds, in the heterotic string frame, to the limit of large volume of the non-perturbative $T^4$ torus. The geometry of $T^3$ in type IIA strings units is also encoded in the moduli $v$. The vielbein $e_{4,20}$ parametrizes the moduli (metric and B-field) of type IIA on K3 and $\psi$ include the Wilson lines of the 6D gauge bosons along $T^3$ as well as the moduli for the 3D scalars dual to these gauge fields.

\bigskip

Let us consider a Niemeier point in the moduli space, where $\Gamma^{8,24} \cong L\oplus E_8(-1)$ and let $H\subseteq \Aut(L)\subset O(8,24,\ZZ)$ be the subgroup of automorphisms of the Niemeier lattice $L$ that fixes $\Gamma^{4,4}$ pointwise. Then, by the same argument as in the previous subsection, any $h\in H$ satisfies
\be h\cdot e_{8,24}=\begin{pmatrix}
\bf{1}_4 & &\\ & h' &\\ && \bf{1}_4
\end{pmatrix}\cdot e_{8,24}=e_{8,24}\cdot \begin{pmatrix}
\bf{1}_4 & &\\ & \rho' &\\ && \bf{1}_4
\end{pmatrix}\ ,
\ee where $h'\in O(\Gamma^{4,20})$ and $\rho'\in SO(20)\subset SO(4)\times SO(20)$. By plugging the expression \eqref{IIAvielbein} into this equation, it is clear that $h$ is a symmetry of the model independently of the volume $\left|\det v\right|^2$, so that the limit $\left|\det v\right|^2\to \infty$ (large volume of the perturbative $T^4$ torus in heterotic string units) is a weakly coupled type IIA string theory on K3$\times T^3$ with discrete symmetry group $H$. One can also deform the geometry of the torus $T^3$ (in particular, decompactify the type IIA theory up to six dimensions) without further breaking the symmetry group $H$.

\bigskip

If we choose $u_1,\ldots,u_4,v_1,\ldots,v_4$ such that no root in $L$ is orthogonal to $\Gamma^{4,4}$, then the weakly coupled model has generic gauge group $U(1)^{30}$ and the perturbative type IIA string is described by a non-linear sigma model on $K3\times T^3$. On the contrary, when the model has enhanced (non-abelian) gauge symmetry, the world-sheet CFT describing the type IIA fundamental string is believed to be singular even for small values of the coupling constant $g_{3IIA}$, due to exactly massless D-brane states.

At the points of generic gauge group, the groups $H$ obtained in this way are therefore symmetries of the corresponding non-linear sigma model on K3. In particular, the $20$-dimensional representation $\rho'\in SO(20)$ is the representation on the $20$ Ramond-Ramond ground states of the model that are singlets under the worldsheet $SU(2)_L\times SU(2)_R$ R-symmetries. Consistently, the groups $H$ are exactly of the form expected from the classification of symmetries of K3 sigma models in \cite{GHV} and \cite{toappear}. In the heterotic frame, the representation $\rho'\in SO(20)$ acts on the $20$ left-moving bosonic oscillators along the internal directions of heterotic on $T^4$. From this construction, it is clear that the symmetry groups $H$ of non-linear sigma models on K3 can get enhanced (at least in some cases) to a full Umbral group or to $Co_0$ when the model is `embedded' in a certain non-perturbative limit of type IIA on K3$\times T^3$.

\subsection{Threshold corrections in perturbative limit}

Let us now analyze the threshold correction \eqref{thresh} to the four scalar coupling in the perturbative heterotic string ($g_{3H}\to 0$). The leading term in this limit is the standard 1-loop threshold correction (which itself was the starting point of \cite{OP} for the conjectural expression \eqref{thresh})
\be g_{3H}^8 \int_{\F} \frac{d^2\tau }{\tau_2^2} p_L^4 Z_{7,23}\frac{1}{\eta^{24}}\ .
\ee The integrand clearly encodes the spectrum of perturbative 1/2 BPS string states in this model: the theta series $Z_{7,23}$ is the sum over the left- and right-moving momenta along the internal Narain lattice $\Gamma^{7,23}$ and $1/\eta^{24}$ is the contribution of the $24$ bosonic (left-moving) oscillators in the transverse directions. The 1/2 BPS condition is implemented by turning off all right-moving transverse oscillators; finally, the contribution of ghosts and superghosts cancels against the longitudinal oscillators, as usual. The level-matching condition for 1/2 BPS states
\be \frac{1}{2}(p_R^2-p_L^2)=N-1\ ,
\ee where $N$ is the level of the bosonic oscillators,
is \emph{not} a priori imposed on the integrand. Rather, it is implemented through the integration over $\tau_1$ (upon unfolding the integral over the fundamental domain $\F$ into an integral over a vertical strip $-1/2<\tau_1<1/2$, $\tau_2>0$). After imposing the level matching condition, the Fourier coefficients of
\be \frac{1}{\eta(\tau)^{24}}=\sum_{n=-1}^\infty d(n)q^n\ ,
\ee are interpreted as the multiplicities $d(n)$ of 1/2 BPS states with electric charge $(p_L,p_R)$, such that $p_R^2-p_L^2=2n$.

In subsection \ref{s:pertHet}, we have seen that there exist perturbative heterotic models with generic gauge group $U(1)^{30}$ and with a symmetry group $G$ equal to any of the Umbral groups (or to any subgroup of $Co_0$ fixing a $1$-dimensional space). Furthermore, we showed that the $23$ bosonic oscillators along the \emph{internal} compactified directions transform in the standard $23$-dimensional representation of $G$, while the transverse oscillator along the uncompactified space-time direction is fixed by $G$. This means that the multiplicities $d(n)$ of 1/2 BPS states naturally decompose into representations of $G$.

\bigskip

Notice that, in general, the group $G$ acts also on the charges $(p_L,p_R)$ -- in fact, it is this action of $G$ that is visible in the decompositions in section \ref{Evaluation}. This means that the symmetry group $G$ maps the space $\Hh^{BPS}_{p_L,p_R}$ of 1/2-BPS perturbative states with fixed charges $p_L,p_R$ to the space $\Hh^{BPS}_{g(p_L,p_R)}$, where $g(p_L,p_R)\in \Gamma^{7,23}$ is a vector of the same norm as $(p_L,p_R)$. Therefore, strictly speaking, there is no well-defined action of $G$ on a single space $\Hh_{p_L,p_R}$, unless $p_L,p_R$ is fixed under $G$. However the full space $\oplus_{p_L,p_R}\Hh_{p_L,p_R}^{BPS}$ has a natural decomposition as (a sum over) tensor products of $G$-representations
\be \bigoplus_{(p_L,p_R)\in \Gamma^{7,23}}\Hh_{p_L,p_R}^{BPS}=\bigoplus_{n\in\ZZ} \bigoplus_{\substack{(p_L,p_R)\in \Gamma^{7,23}\\ p_R^2-p_L^2=2n} }\Hh_{p_L,p_R}^{BPS}=\bigoplus_{n\in\ZZ}(\Hh^{mom}_n\otimes \Hh^{osc}_n)\ ,
\ee where $\Hh^{mom}_n$ is the $G$-representation spanned by vectors of norm $2n$ in $\Gamma^{7,23}$ and $\Hh^{osc}_n$ is the $G$-representation of dimension $d(n)$ induced by the action over the bosonic oscillators.

\section{Discussion}

The 3D picture of the Umbral symmetry groups we have advocated here fits naturally into the developing understanding of the Mathieu and Umbral moonshines.  However, many questions require further exploration.

\medskip
\noindent
$\bullet$ It has been advocated that all 23 Niemeier lattices play an important role in the geometry of K3
surfaces, most concretely via Nikulin's geometry of `Niemeier markings' \cite{Nikulin}.  In IIA string theory
on K3, then, it is natural to hope that symmetry groups at a given point in moduli space can be naturally
associated with automorphisms of a Niemeier lattice marking the given K3 surface \cite{TW} (though the concept must be
suitably extended to the full moduli space of conformal field theories, or even string theories, on K3). Results in this direction will appear in \cite{toappear}.  Our
3D starting point of a set of compactifications which enjoy the automorphisms of each Niemeier lattice as a
symmetry, together with the precise way in which one can track decompactification limits as described briefly
in \S4, suggests that one should be able to derive the appropriate marking of symmetries directly from 3D.

\medskip
\noindent
$\bullet$ The elliptic genus of K3 was the starting point for investigation of new moonshines.  Mathieu moonshine
was discovered there \cite{EOT}, and the full set of Umbral mock modular forms have a suggestive relationship with
the K3 elliptic genus \cite{CH}.  Can we recover facts about the elliptic genus, and the associated mock modular
forms which appear in its character expansion, from the 3D perspective?  This suggests two avenues of investigation:
direct study of spectra of space-time BPS states, and generalization of the elliptic genus to a more refined index
(which absorbs the torus zero modes) which may be non-trivial on $K3 \times T^3$, and which could include the data
stored in the K3 elliptic genus.  Such a refinement may be readily available by generalizing \cite{Greg}.

\medskip
\noindent
$\bullet$ We have focused on a three-dimensional description of gravity theories with Umbral symmetry.  Further compactification to 2D would allow us to discuss Niemeier compactification of the heterotic string (dual to type II on $K3 \times T^4$), where now the lattices we discuss would
show up literally on the worldsheet, instead of capturing the more abstract geometry of moduli in 3D.  This picture
should have some advantages -- for instance, the standard techniques of vertex operator algebra, so prominent in earlier studies of moonshine \cite{FLM, DGH}, will apply. In particular, the constructions in \cite{PV} might generalize to these cases.  From the point of view of explaining symmetries, it has one drawback -- one has ``used up" more dimensions to get to the same global symmetry, so in this sense our 3D story is stronger, requiring less specialization to get back to 6D compactifications.

\medskip
\noindent
$\bullet$ The best known component of the moduli space of theories with 16 supercharges -- that which holds heterotic strings on $T^7$ and type II strings on $K3 \times T^3$ -- has yielded a rich
story involving algebraic geometry, lattice theory, number theory, and so forth.  There are other components
of the moduli of theories with 16 supercharges, however (see e.g. \cite{components} for some discussion of this fact).
Could equally rich stories be hiding in the other components?

\medskip
\noindent
$\bullet$ The story we have told could be discussed in the duality frame of M-theory on $K3 \times T^4$, where it
becomes entirely geometric (in the sense that no B-fields or worldsheet quantities are involved in the description,
though preferred points in moduli space can have ${\cal O}(1)$ volumes in Planck units).  This suggests that there
may be interesting viewpoints on Mathieu, Umbral and other moonshines hidden in the geometry of 8-manifolds.

\bigskip
\noindent
{\centerline{\bf{Acknowledgements}}

\medskip
\noindent
We would like to thank M. Cheng and S. Harrison for relevant helpful discussions over the past few years. R.V. would like to thank Boris Pioline for useful discussions about the special points in the moduli space of 3D heterotic strings. S.K. is supported by the National Science Foundation
under grant NSF-PHY-1316699. N.M.P. is supported by a National Science Foundation Graduate Research Fellowship under grant DGE-114747.

\appendix
%

\section{Evaluating the integral}\label{app}
We can readily compute the one-loop integral in the main text using the techniques developed in \cite{Pioline}, whose treatment we follow. To be precise, we are faced with a modular integral of the form
\begin{equation}
\int_{\mathcal{F}}d\mu \left(\tau_2^{-\lambda/2}\sum_{p_L, p_R}\rho(p_L \sqrt{\tau_2}, p_R \sqrt{\tau_2})q^{p_L^2/2}\bar{q}^{p_R^2/2} \right) \Phi(\tau).
\end{equation}
The integral is over the fundamental domain, $d\mu = d^2 \tau/(\tau_2)^2$ is the modular-invariant measure, the sum is over a lattice of signature $(d, d+k) = (8,24)$, and $\Phi(\tau)$ is, in general, an almost-holomorphic modular form of weight $w$ (i.e. it is an element of the graded ring generated by $E_4, E_6, 1/\Delta, \hat{E_2}$). The term in parenthesis is a modular form of weight $(\lambda + d + k/2, 0)$ and so the integrand is modular invariant as required if we impose $\lambda + d + k/2 = -w$. Moreover, the polynomial $\rho(x_L, x_R)$ must be annihilated by the modular covariant derivative
\begin{equation}
\sum_{i=1}^{d+k} \partial^2_{x_{L, i}} - \sum_{i=1}^{d} \partial^2_{x_{R, i}} - 4 \pi (\sum_{j=1}^{d+k}x_{L, j} \partial_{x_{L, j}} -  \sum_{j=1}^{d} x_{R, j} \partial_{x_{R, j}} - \lambda - d) \rho(x_L, x_R)=0.
\end{equation}
Here, $x_{L/R, i}$ denotes the $i$th component of the $(d+k)$- (respectively, $d$-) dimensional momentum vector, multiplied by $\sqrt{\tau_2}$. One can check that the covariantized four-momentum insertion $x_{L, 1}^4 - \frac{3}{2 \pi}x_{L, 1}^2 + \frac{3}{16 \pi^2} = \tau_2^2 \left(p_{L, 1}^4 - \frac{3}{2 \pi \tau_2}p_{L, 1}^2 + \frac{3}{16 \pi^2 \tau_2^2} \right)$ is annihilated by this operator if $\lambda = 4 - d$, where we have chosen the momenta to all point in the $1$-direction. Notice that this gives an additional overall power of $\tau_2^2$, which includes the usual contribution $\tau_2^{d/2}$ from the lattice.

Incorporating the results of \cite{Pioline}, this means we can instead evaluate an integral of the form
\begin{equation}\label{eq:int1}
\int_{\mathcal{F}}d\mu \tau_2^{-\lambda/2} \sum_{p_L, p_R} \rho(p_L \sqrt{\tau_2}, p_R \sqrt{\tau_2})q^{p_L^2/2}\bar{q}^{p_R^2/2} \mathcal{F}(s, \kappa, w)
\end{equation}
where $\mathcal{F}(s, \kappa, w)$ is a Niebur-Poincare series, $\kappa$ is the width of the cusp at $\tau \rightarrow i\infty$, $w$ is the weight, and $s$ is an analytic continuation parameter. For suitable values of $s$, the Niebur-Poincare series recovers the original $\Phi(\tau)$ \footnote{In general, one must take a linear combination of Niebur-Poincare series to recover one's original form, but our case is particularly simple.}. The benefit of using a Niebur-Poincare series representation is that it provides a natural regularization of the Poincare series for a negative weight (weak) almost holomorphic modular form, and moreover has the nice property that it is annihilated by the Laplacian on the upper half plane. Explicitly, one can write the series as 
\begin{equation}
\mathcal{F}(s, \kappa, w) = \frac{1}{2}\sum_{\gamma = \Gamma_{\infty} \backslash \Gamma} \mathcal{M}_{s, w}(-\kappa \tau_2) e^{- 2 \pi i \kappa \tau_1} \vert_{w, \gamma}
\end{equation}
and $\mathcal{M}_{s, w}(y)$ is the Whittaker function.

In particular, if the weight is negative, $w <0$, one is in the region of absolute convergence of this series and can analytically continue the series to the value $s = 1- w/2$, where it becomes precisely the (weak) holomorphic modular form of interest. Namely, for our purposes $\Phi(\tau) = \frac{1}{\Delta(\tau)}$ (see appendix \ref{a124example}), with $\kappa=1, w= -12$.  With this specialization one has, in the $s \rightarrow 1- w/2 = 7$ limit, $\mathcal{F}(1 - w/2, 1, w) \rightarrow \mathcal{F}(7, 1, -12) = 13!/\Delta $. 

The authors of \cite{Pioline} then show that given an integral in the form of \ref{eq:int1}, if $\rho$ is a polynomial in $p_L^a, p_R^b$, can be evaluated term-by-term as \footnote{The extra factors of $\sqrt{2}$ come from a rescaling of the momenta in our conventions relative to those of \cite{Pioline}.}
\begin{align}\nonumber
\int_{\mathcal{F}} d\mu \tau_2^{\delta} (\sqrt{2})^{\sum_{i=1}^{\alpha} a_i + \sum_{i=1}^{\beta} b_i} &\sum_{p_L, p_R}p_L^{a_1} \ldots p_L^{a_{\alpha}} p_R^{b_1} \ldots p_R^{b_{\beta}} q^{p_L^2/2}\bar{q}^{p_R^2/2} \mathcal{F}(s, \kappa, w) \\ \nonumber
&= (4 \pi \kappa)^{1 - \delta} (\sqrt{2})^{\sum_{i=1}^{\alpha} a_i + \sum_{i=1}^{\beta} b_i} \Gamma(s + |w|/2 + \delta-1) \\ \nonumber &\times \sum_{p_R^2 - p_L^2 = -2}p_L^{a_1} \ldots p_L^{a_{\alpha}} p_R^{b_1} \ldots p_R^{b_{\beta}} {}_2 F_{1}\left(s - |w|/2, s + |w|/2 + \delta -1, 2 s, \frac{2 \kappa}{p_L^2} \right) \\ \nonumber &\times(p_L^2/(2 \kappa))^{1 - s - |w|/2 - \delta}
\end{align}
where $\delta = (\alpha + \beta + \delta)/2$. In our particular case, we will fix $\kappa=1$ and divide by $13!$ per the relation between the Niebur-Poincare series and $1/\Delta$. We will always have $\beta=0$, and three terms with $\alpha= 4, 2, 0$ and $\delta = 4, 3, 2$, respectively. We will also have the constants $1, 3/(2\pi), 3/(16 \pi^2)$, fixed by the modular derivative, multiplying the three terms.

If we specialize all of these quantities then the answer is

\begin{align}\nonumber
\frac{4}{13!} \frac{1}{(4 \pi)^3}\sum_{p_R^2 - p_L^2 = -2} & \Gamma(16) p_{L, 1}^{4} { \ }_2 F_{1}\left(1, 16, 14, \frac{2}{p_L^2} \right) (p_L^2/2)^{-16} \\ \nonumber & + 6 \Gamma(15) p_{L, 1}^{2} { \ }_2 F_{1}\left(1, 15, 14, \frac{2}{p_L^2} \right) (p_L^2/2)^{-15} 
\\ \nonumber &+  3 \Gamma(14) { \ }_2 F_{1}\left(1, 14, 14, \frac{2}{p_L^2} \right) (p_L^2/2)^{-14} 
\end{align}
or, equivalently, the summand becomes
\begin{align*}\nonumber
&\frac{2981888 p_{L, 1}^4}{\pi ^3 \left(1-\frac{2}{p_{L}^2}\right)^3 p_{L}^{36}}-\frac{3194880 p_{L, 1}^4}{\pi ^3
   \left(1-\frac{2}{p_L^2}\right)^3 p_L^{34}}+\frac{860160 p_{L, 1}^4}{\pi ^3 \left(1-\frac{2}{p_L^2}\right)^3
   p_L^{32}} \\& \nonumber -\frac{319488 p_{L, 1}^2}{\pi ^3 \left(1-\frac{2}{p_L^2}\right)^2 p_L^{32}}+\frac{172032 p_{L, 1}^2}{\pi ^3
   \left(1-\frac{2}{p_L^2}\right)^2 p_L^{30}}+\frac{3072}{\pi ^3 \left(1-\frac{2}{p_L^2}\right) p_L^{28}}
\end{align*}

\section{The theta series of the Niemeier lattice $A_1^{24}$}\label{a124example}

As an example of the construction in section \ref{Evaluation}, let us take $L^X$ to be the Niemeier lattice with root lattice $X=A_1^{24}$ and $\lambda$ is the Weyl vector corresponding to a choice of positive roots. The theta series $\Theta_{L(A_1^{24}),\lambda}$ is a Jacobi form of weight $12$ and index $\lambda^2/2=6$, which can be found by expanding on a standard basis of Jacobi forms. The final result is
\begin{align*}
\Theta_{L(A_1^{24}),\lambda}=&
\frac{1}{2985984}B^6 E_4^3-\frac{1}{497664} B^5AE_4^2E_6+  \frac{5}{995328} B^4A^2E_4^4-\frac{5}{746496}B^3A^3E_4^3E_6\\
&+\frac{5}{99
   5328} B^2A^4E_4^5-\frac{1}{497664}BA^5E_4^4E_6+\frac{1}{2985984}A^6E_4^6-\frac{7}{31104}B^6\Delta\\&-\frac{29}{5184}B^4A^2\Delta E_4+\frac{29}{
   7776}B^3A^3\Delta E_6-\frac{31}{3456}B^2A^4\Delta E_4^2\\&+\frac{7}{2592}BA^5\Delta E_4E_6-\frac{7}{7776}A^6\Delta E_4^3+\frac{14}{9}A^6\Delta^2\\
=& 1+q \left(24
   y+\frac{24}{y}\right)
   +q^2 \left(759 y^4+\frac{759}{y^4}+6072 y^3+\frac{6072}{y^3}+21528
   y^2+\frac{21528}{y^2}\right.
\\&\qquad\qquad\qquad\qquad\qquad\left.+42504 y+\frac{42504}{y}+53682\right)+\ldots
\end{align*} 
Here, $A$ and $B$ are the standard weak Jacobi forms of index $1$ and weights, respectively, $-2$ and $0$ (see \cite{EichlerZagier})
\be A=\phi_{-2,1}(\tau,z)=\frac{\vartheta_1(\tau,z)^2}{\eta(\tau)^6}=\frac{1}{y}-2+y+\ldots
\ee
\be B=\phi_{0,1}(\tau,z)=4\sum_{i=2}^4\frac{\vartheta_i(\tau,z)^2}{\vartheta_i(\tau,0)^2}=\frac{1}{y}+10+y+\ldots~,
\ee with $\vartheta_i$ the standard Jacobi theta series, $E_k$ are the Eisenstein series of weight $k$
\be E_k(\tau)=1-\frac{2k}{B_{2k}}\sum_{n=1}^\infty \sigma_{k-1}(n) q^n\ ,
\ee with $B_{2k}$ the Bernoulli numbers and $\sigma_{k-1}(n)=\sum_{d|n}d^{k-1}$. Finally, $\Delta$ is the cusp form of weight $12$
\be \Delta(\tau) = \eta(\tau)^{24}=q\prod_{n=1}^\infty (1-q^n)^{24}\ .
\ee
The Fourier coefficients of $\Theta_{L(A_1^{24}),\lambda}$ decompose naturally into representations of the associated Umbral group  $G^X=M_{24}$
\begin{align*}
c(1,1)&=24={\bf 1}+{\bf 23}\\
c(2,4)&=759={\bf 1} + {\bf 23} + {\bf 252} + {\bf 483}\\ 
c(2,3)&=6072= {\bf 1} + 2\times {\bf 23} + 2\times {\bf 252} +{\bf 253}  + {\bf 483}+{\bf 1265}+{\bf 3520} \\
&\vdots
\end{align*}

\end{document}